# Efficient and Stable PbS Quantum Dot Solar Cells by Triple-Cation Perovskite Passivation

*Miguel Albaladejo-Siguan[1,2], David Becker-Koch[1,2], Alexander D. Taylor[1,2], Qing Sun[1], Vincent Lami[1], Pola Goldberg Oppenheimer[3], Fabian Paulus[1,2] and Yana Vaynzof\*[1,2]*

[1]Kirchhoff Institute for Physics, Heidelberg University, Im Neuenheimer Feld 227, 69120 Heidelberg, Germany

[2]Integrated Centre for Applied Physics and Photonic Materials and Centre for Advancing Electronics Dresden (cfaed), Technical University of Dresden, Nöthnitzer Straße 61, 01187 Dresden, Germany

[3]School of Biochemical Engineering, University of Birmingham, Edgbaston, Birmingham, West Midlands B15 2TT, United Kingdom

ABSTRACT

Solution-processed quantum dots (QDs) have a high potential for fabricating low cost, flexible and large-scale solar energy harvesting devices. It has recently been demonstrated that hybrid devices employing a single monovalent cation perovskite solution for PbS QD surface passivation exhibit enhanced photovoltaic performance when compared to standard ligand passivation. Herein we demonstrate that the use of a triple cation $Cs_{0.05}(MA_{0.17}FA_{0.83})_{0.95}Pb(I_{0.9}Br_{0.1})_3$ perovskite composition for surface passivation of the quantum dots results in highly efficient solar cells, which maintain 96 % of their initial performance after 1200h shelf storage. We confirm perovskite shell formation around the PbS nanocrystals by a range of spectroscopic techniques as well as high-resolution transmission electron microscopy. We find that the triple cation shell results in a favorable energetic alignment to the core of the dot, resulting in reduced recombination due to charge confinement without limiting transport in the active layer. Consequently, photovoltaic devices fabricated *via* a single-step film deposition reached a maximum AM1.5G power conversion efficiency of 11.3 % surpassing most previous reports of PbS solar cells employing perovskite passivation.

Colloidal quantum dots (QDs) of semiconducting materials are a promising source for fabricating efficient photovoltaic devices offering low-cost ink processing, increased stability, and size-controlled bandgap tunability.[1–4] In the last few years, advances regarding the passivation of surface states,[5,6] as well as device optimization[7–9] have led to a certified efficiency record of 16.6 %.[10] A common QD material is lead sulfide (PbS), which absorbs in the visible-IR range of the solar spectrum, and can be either n- or p-doped by exchanging the oleic acid (used for solution stability) with a different ligand.[11] The size of the QDs (3-4 nm) determines the width of the bandgap (1.3-1.0 eV), which correlates with the open circuit voltage in the corresponding photovoltaic devices.[12]

At the same time, solar cells utilizing organic-inorganic hybrid perovskite (PVK) materials have reached efficiencies around 23 % with an optimized active layer composition based on a triple cation structure $Cs_{0.05}(MA_{0.17}FA_{0.83})_{0.95}Pb(I_{0.9}Br_{0.1})_3$ (abbreviated as CsMAFA, which stands for cesium, methylammonium and formamidinium).[13,14] The combination of both material systems into a single device offers intriguing prospects, with the potential to combine the advantageous properties of both: the high open circuit voltage and fill factor in the case of perovskite solar cells, and the increased short circuit current due to the infrared absorption for QD solar cells. Recent reports have demonstrated such an approach, for example by incorporating QDs between PVK domains,[15] using QDs as a thin modifying interlayer,[16] or by growing a perovskite shell on the surface of QDs.[17–19] In the case of PbS QDs, the lattice constant is one of the limiting factors for choosing an appropriate PVK shell, since a large lattice mismatch would induce an undesired strain.[20] So far, it has been demonstrated that single monovalent cation perovskites, such as $MAPbI_3$, can be successfully grown on the surface of PbS QDs[17,18,21] as well as $CsPbI_3$-PVK, or other metal-halides with the perovskite structure.[19,20] In both cases, solution-based ligand exchange simplifies the process for device fabrication, and minimizes the use of solvents as compared to layer-by-layer spin-coating.[22,23] The choice of cation influences the performance of the device, with MAPbI$_3$-PbS cells reaching 9% power conversion efficiency,[18] while the highest efficiencies of CsPbI$_3$-PbS solar cells are around 10.5 %.[19] Liu *et al.* have recently demonstrated that it is also possible to embed PbS QDs in a CsPbBrI$_2$ matrix (~15 % PVK content) and by that fabricate solar cells with 12.6 % PCE.[24] At the same time, other mixed organic/inorganic ligands (*e.g.* MAI or PbI$_2$/PbBr$_2$) have been used to build highly efficent solar cells, reaching record efficencies of 12 %.[22,25]

In this study, we report the formation of a triple cation-double anion perovskite shell around PbS quantum dots, and demonstrate their application in efficient and stable photovoltaic devices. The perovskite crystal formation, as well as the increased interdot coupling and surface passivation, are confirmed by spectroscopic analysis using X-ray photoemission spectroscopy (XPS) and Fourier-transform infrared spectroscopy (FT-IR) as well as by high-resolution transmission electron microscopy imaging. Devices based on CsMAFA-PbS active layers result in excellent photovoltaic performance with average efficiencies of 10.6% and a record efficiency of 11.3%, surpassing previous reports on PVK-PbS devices. We demonstrate that this excellent performance is associated with the energetic alignment between the CsMAFA shell and the PbS core that results in reduced recombination due to improved charge confinement, without adversely affecting charge transport. Finally, the devices maintain 96 % of their initial PCE after 1200h of shelf storage, making triple cation perovskite coating a promising approach for the fabrication of efficient and stable photovoltaic devices.

RESULTS AND DISSCUSSION

**Ligand exchange and perovskite shell formation.** Oleic acid covered PbS QDs were synthesized *via* a hot injection method following a previously published recipe.[26] The ligand exchange (**Figure 1a**) was performed in a two-phase solution, using the iodide salts of Cs, MA and FA combined with $PbX_2$ (X = I, Br) in DMF/DMSO (0.2 mol/L) as the perovskite precursor and a 10 mg/ml OA-PbS QD solution. After three solvent washing steps, the QDs were precipitated by addition of toluene, and collected by centrifugation. A highly concentrated CsMAFA-PbS ink (300 mg/ml in butylamine) was prepared for single-step film deposition followed by a 30 min annealing step at 100 °C. The reference samples were prepared using similar fabrication techniques adapted from literature,[21,23] with either an inorganic $PbX_2$ (X = I, Br) or $MAPbI_3$ perovskite coating of the PbS QDs.

Following ligand exchange and film deposition, the optical properties of the QD films were measured *via* UV-vis absorption spectroscopy. As can be seen in **Figure 1b**, the absorbance in the spectral range of 400 to 1100 nm is similar for both the CsMAFA-PbS and $PbX_2$-PbS films and is in agreement with the characteristic absorbance of PbS films with lead halides as ligands.[17] **Figure 1c** shows the absorption coefficients of the same films, which shows no spectral features that can be associated with separate perovskite domain formation, such as absorption increase below ~800 nm. Photoluminescence (PL) and absorbance spectra measured on highly concentrated solutions of the triple cation capped QDs reveal a 15 meV redshift in the peak position as compared to the fully inorganic $PbX_2$-PbS reference (**Figure 1d**), which is associated with a higher dot-to-dot interaction.[17,27] The same shift is also observed in thin films (Figure 1b). The Stokes shift between absorbance and emission is 140 meV for both CsMAFA and $PbX_2$, which is consistent with the typical Stokes shift for PbS QDs reported in previous works.[28,29]

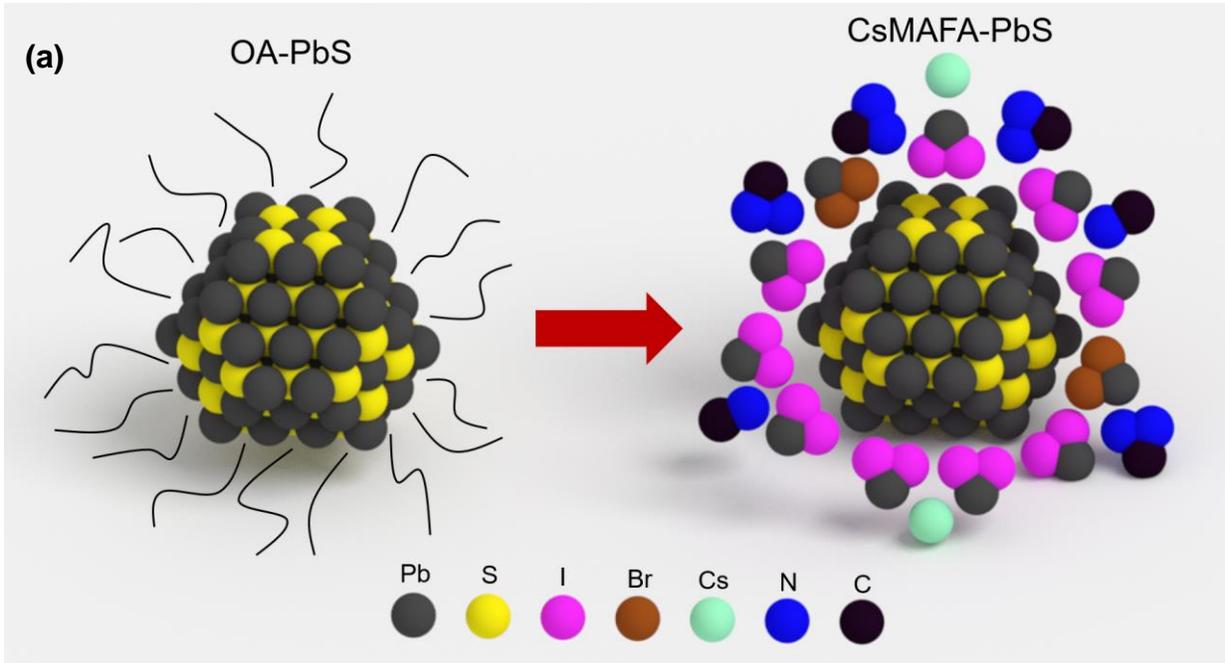

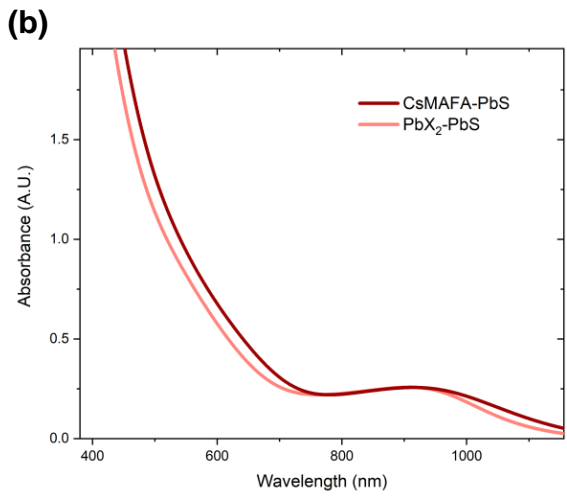
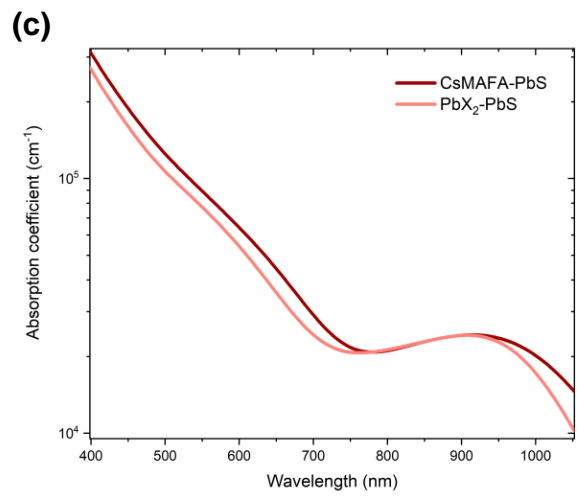
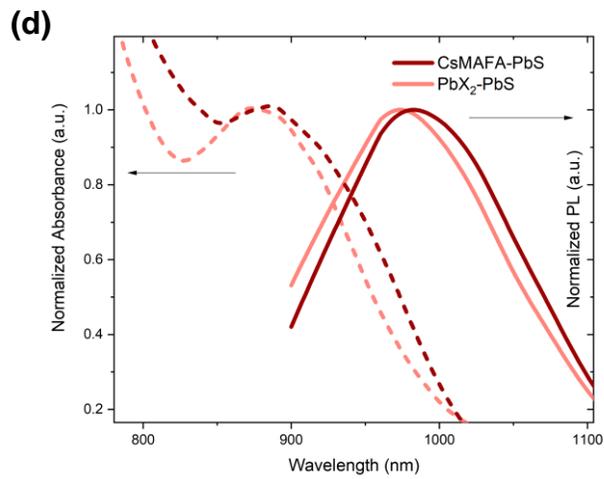

**Figure 1.** Surface treatment of OA-PbS QDs. a) Schematic representing the solution ligand exchange process used in this study. b) UV-vis absorption of QD films on glass. c) Absorption coefficient of the QD films and d) normalized PL emission and absorbance of the QD solution.

**Characterization of the QDs after passivation.** To confirm the removal of oleic acid on the PbS QD surface and further investigate the formation of a thin perovskite shell, a range of spectroscopic techniques were employed. FTIR measurements (**Figure 2a**) show that the CsMAFA-PbS films exhibit the characteristic vibrations of methylammonium ($MA^+$) and formamidinium ($FA^+$) cations after ligand exchange and washing. The N-$H_x$ stretch originating from $MA^+$ is located at 3500-3100 $cm^{-1}$ while the sharp C=N stretch vibration at 1700 $cm^{-1}$ corresponds to the $FA^+$ cation. We also observe C-$H_x$ stretch vibrations between 2950-2850 $cm^{-1}$ for dots shelled with OA, which are suppressed in the $PbX_2$ reference QD solid film, indicating successful ligand exchange. These C-$H_x$ stretch vibrations remain present for the perovskite coated dots and stem from the $MA^+$ and $FA^+$ cations.[30] The OA removal is also confirmed by the suppression of the peak at 3006 $cm^{-1}$ upon ligand exchange, as is reported also in other publications.[31]

X-ray diffraction measurements confirm that upon addition of Cs, MA, and FA iodide the films do not exhibit separate domains of pure triple cation perovskite. The most characteristic XRD peak for 3D crystalline perovskite occurring at 2θ of ~14° is absent, further supporting the conclusion that no distinct perovskite phase is formed.[32] The diffractograms of both $PbX_2$-PbS and CsMAFA-PbS films exhibit broad reflexes that exclusively originate from the PbS lattice and are indexed in **Figure 2b**. In both cases, the very thin shell of $PbX_2$ and triple cation perovskite could not be resolved in the XRD measurements, similar to previous reports for single cation perovskite coatings.[18]

Finally, XPS spectra (**Figure 2c-g**) of CsMAFA passivated QDs show the presence of a Cs 3$d$ doublet (725 eV and 739 eV) as well as an N 1$s$ singlet (400.8 eV), which further confirm the successful formation of a perovskite shell. The peak positions are comparable with those measured for pure triple cation films (**Figure S1**). The Pb 4$f$ and I 3$d$ peaks are found at similar binding energies as in the reference PbX$_2$ sample. We note that, as expected, Br is present in both the PbX$_2$-PbS and CsMAFA-PbS films.

In order to visualize the triple cation perovskite shell on the QDs, we acquired high-angle annular dark-field scanning TEM (HAADF-STEM) images of the QD after ligand exchange. **Figure 2h** shows the ~2 nm thin perovskite shell on a PbS QD, where the lattice spacings of both cubic PbS and CsMAFA can be identified, as well as their relative orientations. The TEM images confirm a successful epitaxial perovskite growth on the PbS QD surface.

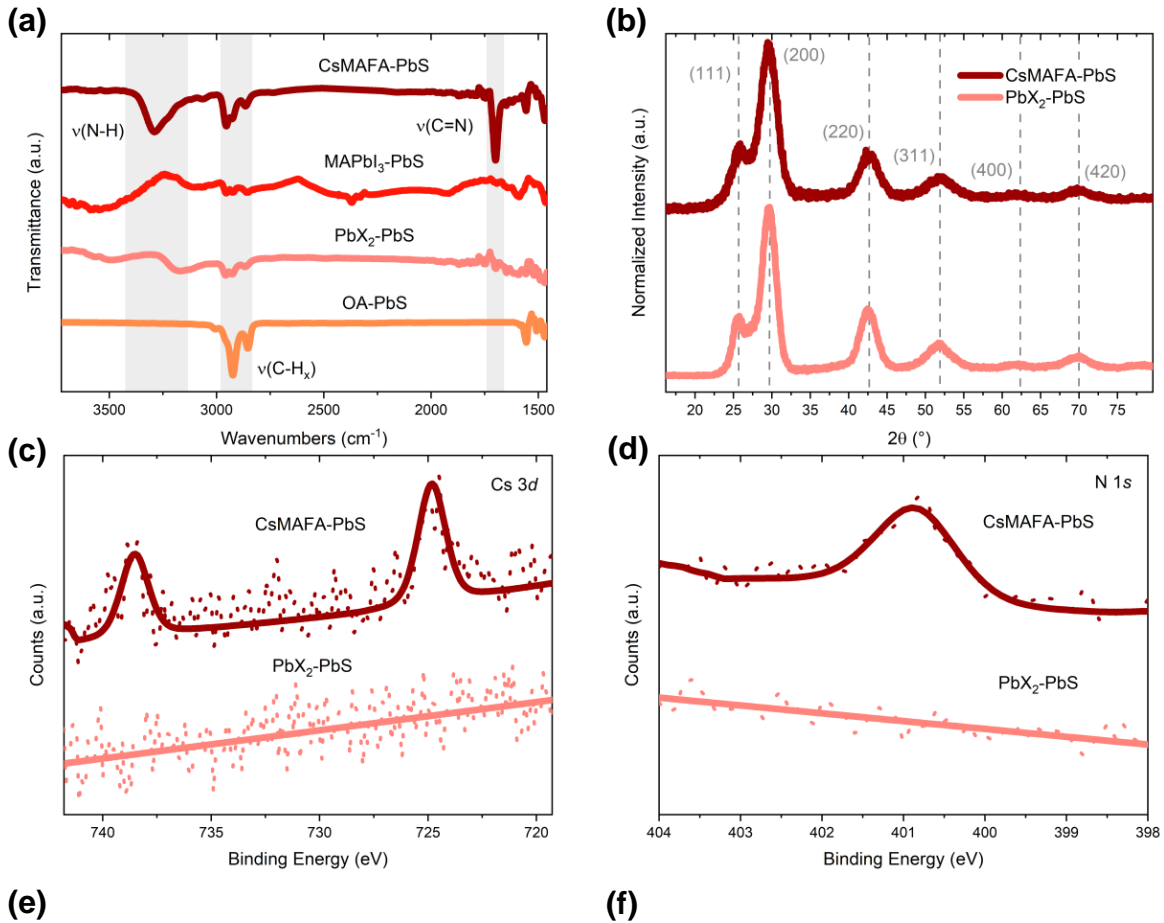

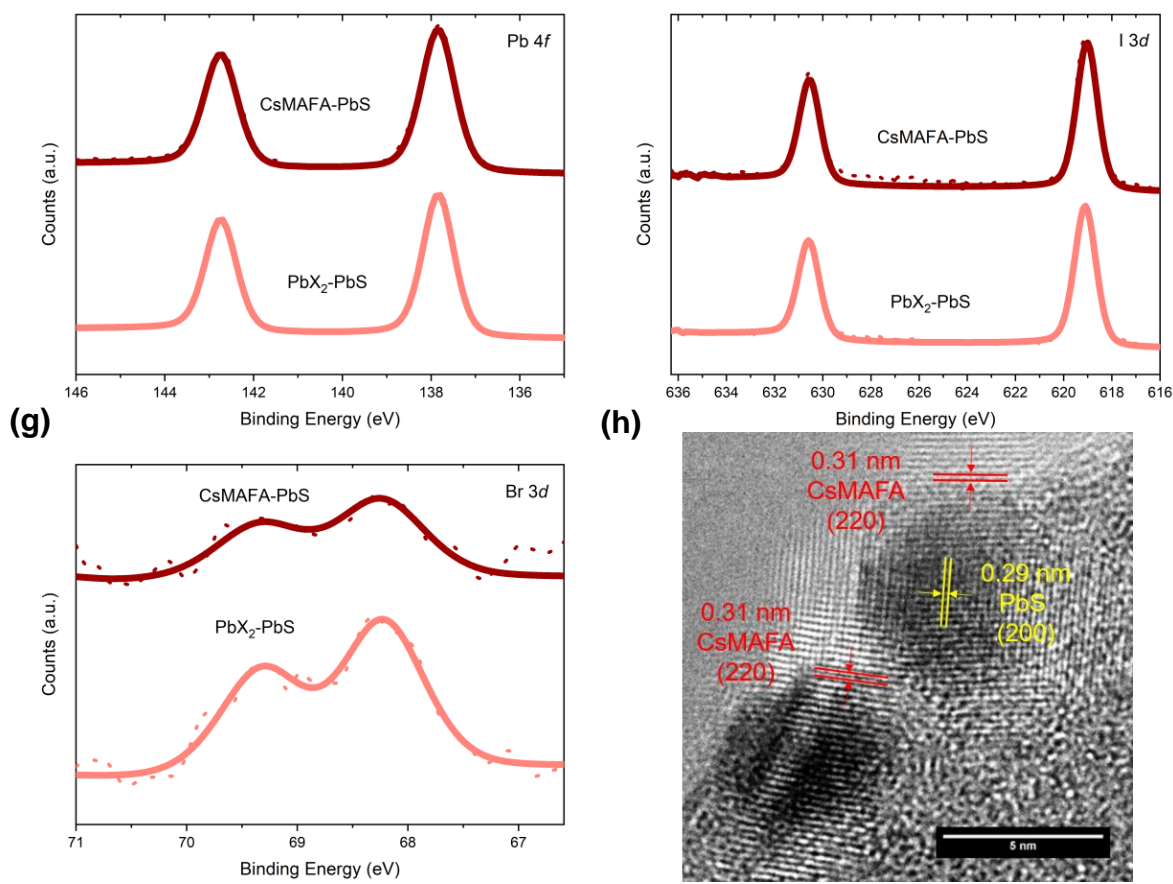

**Figure 2.** Thin film characterization of CsMAFA-PbS QDs. (a) Fourier transform infrared (FT-IR) spectra of PbS QDs before and after ligand exchange. The absorption bands for molecular vibrations of present functional groups are highlighted in grey. (b) X-ray diffraction (XRD) spectra of reference (PbX$_2$-PbS) and CsMAFA-PbS films on glass. X-ray photoemission spectroscopy (XPS) results of reference and CsMAFA-PbS films, showing (c) Cs 3*d* (d) N 1*s* (e) Pb 4*f* and (f) I 3*d*. (g) Br *3d* (h) STEM-HAADF images of CsMAFA-PbS QDs.

**Photovoltaic device performance characterization.** To investigate the effect of CsMAFA coating on the performance of PbS solar cells, devices with the structure Glass/ITO/ZnO/PbS/EDT-PbS/Au (EDT= 1,2-Ethanedithiol) were fabricated and characterized. In short, pre-patterned ITO/Glass substrates were coated by a ZnO sol-gel layer ,[33,34] and the active layer was subsequently formed by single-step deposition of the CsMAFA-PbS QDs ink, followed by a 30 min annealing step at 100°C in air to complete the perovskite formation. The p-type hole extraction layer (EDT-PbS) was deposited *via* solid-state ligand exchange, followed by a 70 nm Au thermally evaporated electrode. The complete device structure can be seen schematically in **Figure 3a**. The stacked device was also imaged by cross sectional scanning electron microscopy (SEM) (**Figure 3b**), confirming the formation of a smooth, uniform CsMAFA-PbS active layer (**Figure S2**), which shows no separate perovskite domains. For comparison, reference devices with either $PbX_2$ or $MAPbI_3$ surface passivation were also fabricated.

The photovoltaic performance measured under 1 sun illumination (AM 1.5G) reveals that CsMAFA-PbS solar cells significantly outperform both the $MAPbI_3$-PbS and $PbX_2$-PbS reference devices. In particular, the short circuit current ($J_{SC}$) and fill factor (FF) are significantly enhanced, resulting in a maximum power conversion efficiency of 11.3 % (**Figure 3c**). The increased $J_{SC}$ can also be directly extracted from the external quantum efficiency measurement, especially in the NIR regime where the EQE peaks at 60 % (**Figure 3d**). We note that the $V_{OC}$ of the devices with all active layers is approximately 0.02 V lower than those reported by Xu *et al* and Zhang *et al*, due to the use of sol-gel derived ZnO instead of ZnO nanoparticles as electron extraction layer, however we find the former to lead to far more reliable and reproducible devices.[19,22] It is also noteworthy that unlike the reference devices, the CsMAFA-PbS solar cells exhibit nearly no hysteresis. Further statistics of the photovoltaic performance can be found in **Table 1**.

To monitor the stability of the devices, we followed the evolution of their photovoltaic performance over time. The unencapsulated devices were transferred from a nitrogen filled glovebox to ambient air for the first measurement and were stored later in air under no illumination for further testing. We observe that the CsMAFA-PbS devices show an enhanced performance after storage in air, similar to other types of PbS devices, such as $PbX_2$-PbS.[7,23] The CsMAFA devices maintain 96 % of their original performance after ~1200 hours (**Figure S3**), suggesting that only very limited degradation takes place. In comparison, the performance of $MAPbI_3$-PbS and $PbX_2$-PbS devices has declined to 85 % and 93 % of their initial PCE after only 480h. In the case of the $MAPbI_3$-PbS devices, the device performance begins to deteriorate directly after fabrication, and are likely associated with the degradation of the $MAPbI_3$ shell, similar to the severe degradation observed in $MAPbI_3$ perovskite based solar cells.[35]

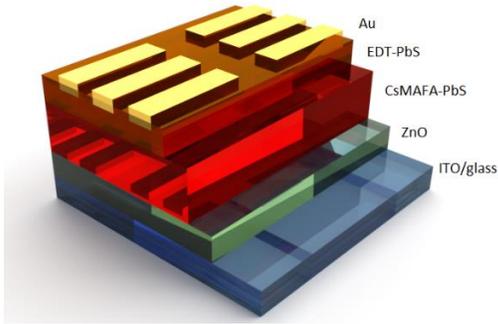
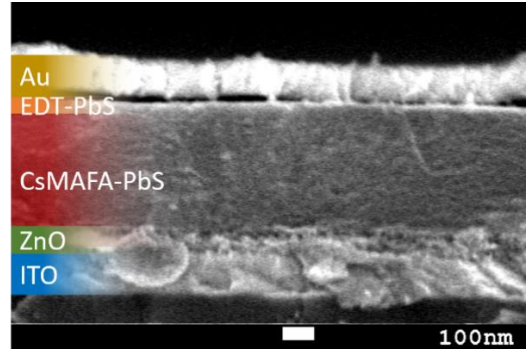
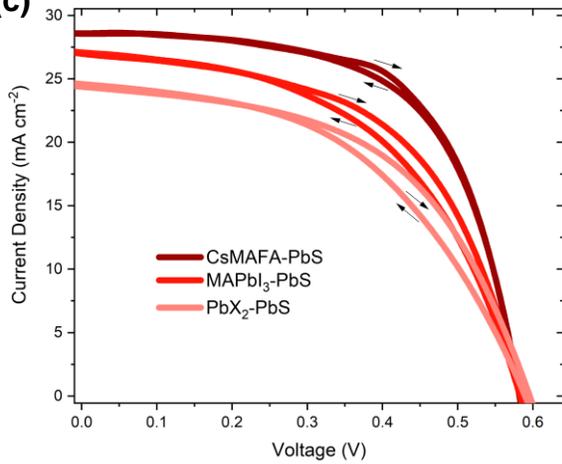
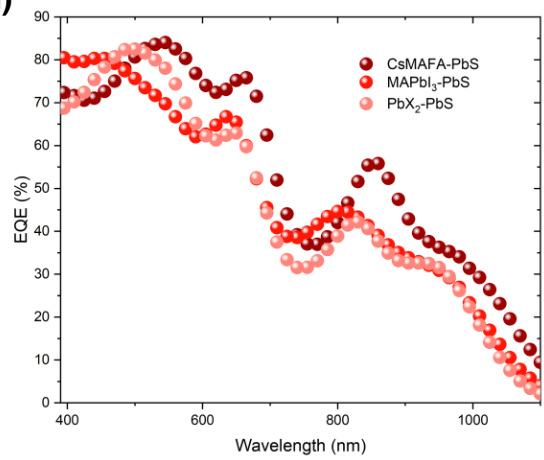

**Figure 3.** Photovoltaic devices made from CsMAFA-PbS QDs. (a) Schematic of the device architecture. (b) Cross-sectional SEM image. (c) Current-voltage scans of (i) CsMAFA-PbS, (ii) MAPbI$_3$-PbS and (iii) PbX$_2$-PbS under AM1.5G illumination. (d) External quantum efficiency (EQE) spectra of CsMAFA-PbS and control samples.

[a)] (Hysteresis index = 1-$PCE_{reverse}$/$PCE_{forward}$)

**Table 1.** Photovoltaic performance of CsMAFA-PbS and control devices. The photovoltaic parameters measured for the champion device of each type (**Figure 3c**) are listed in parentheses.

| QD type | $V_{OC}$ [V] | $J_{SC}$ [mA cm$^{-2}$] | FF [%] | PCE [%] | Hysteresis index [%][a)] |
|---|---|---|---|---|---|
| CsMAFA-PbS | 0.59 ± 0.01 (0.59) | 27.4 ± 1.3 (28.9) | 65.6 ± 1.1 (66.4) | 10.6 ± 0.4 (11.3) | 2 ± 0.2 (1) |
| MAPbI$_3$-PbS | 0.61 ± 0.01 (0.61) | 26.1 ± 0.8 (27.1) | 62.0 ± 2.1 (62.8) | 9.9 ± 0.3 (10.3) | 4 ± 0.5 (5) |
| PbX$_2$-PbS | 0.61 ± 0.01 (0.61) | 23.4 ± 1.2 (24.6) | 53.0 ± 2.0 (56.3) | 7.5 ± 0.2 (8.4) | 8 ± 0.8 (9) |

**Active layer thickness optimization.** In order to gain further insights into the origin of the improved current generation in the hybrid triple cation devices, we performed thickness dependent measurements of the photovoltaic performance. We varied the active layer thickness in the solar cells from 275 nm to 375 nm, in steps of 25 nm, while the bandgap of the QD active layer was kept constant. The results show an optimal thickness at around 350 nm, mainly due to the enhanced $J_{SC}$ (**Figure 4b**) and FF (**Figure 4c**). A further increase of the thickness drastically reduces the charge extraction, meaning that the charge carriers recombine prior to extraction due to the longer distance they must overcome before reaching the extraction layers. We note that the open-circuit voltage (**Figure 4a**) shows only a minor decrease for increased active layer thickness which originates from a higher recombination,[36] yet does not strongly influence the optimal thickness for the maximum PCE (**Figure 4d**). It is interesting to compare the optimal thickness of the CsMAFA-PbS devices to that of the reference devices: 250 nm and 325 nm for MAPbI$_3$-PbS and PbX$_2$-PbS devices, respectively.[21,23] Yang *et al* reported that the short carrier diffusion length within the MAPbI$_3$-PbS active layer is responsible for the decrease in photovoltaic performance for thicknesses above 250 nm.[21] The fact that CsMAFA-PbS devices exhibit a significantly higher optimal thickness suggests that the carrier diffusion length is increased in these devices, warranting a more detailed investigation into the charge carrier dynamics and recombination in these cells.

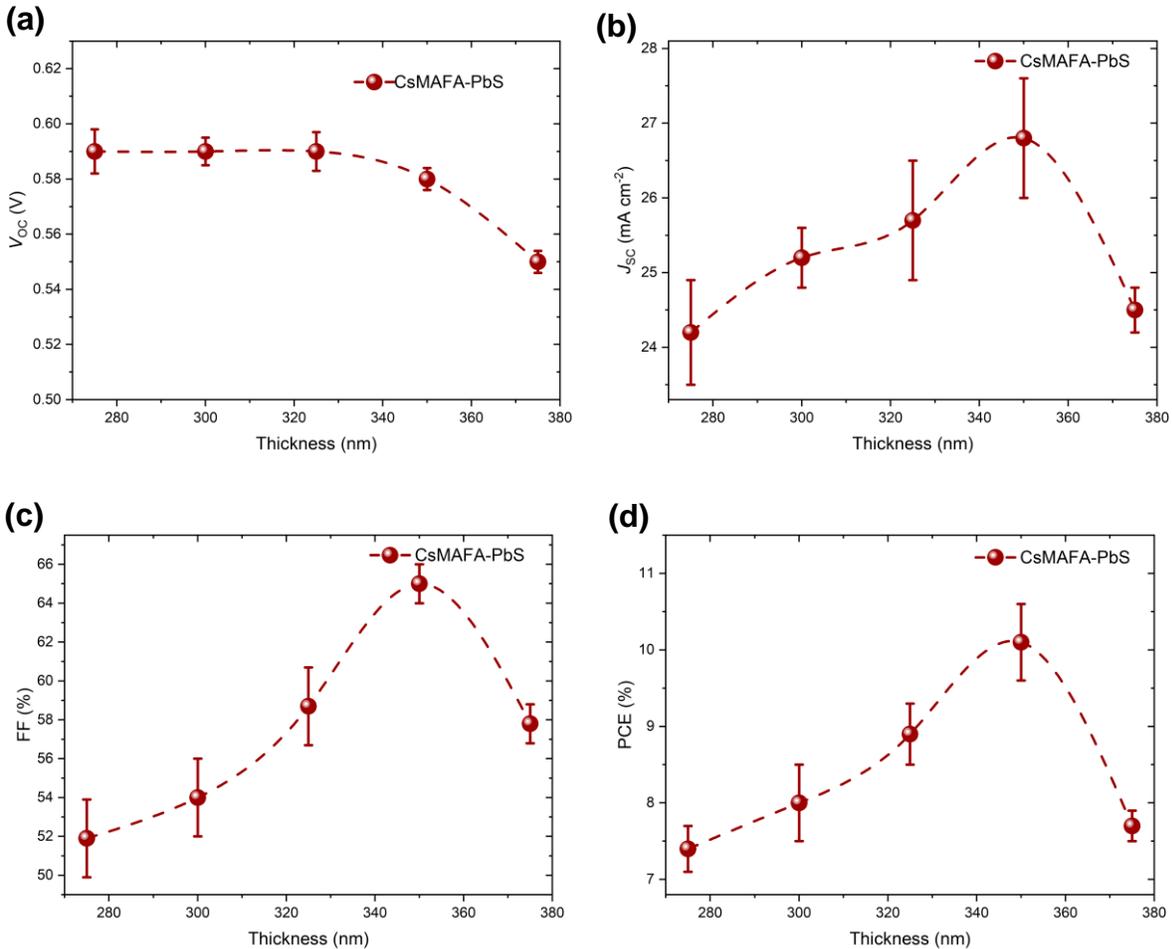

**Figure 4.** Active layer thickness optimization. Thickness dependence of the (a) $V_{OC}$, (b) $J_{SC}$, (c) FF, and (d) PCE for CsMAFA-PbS devices. The error bars represent the standard deviation of 10 devices for each thickness.

**Charge carrier dynamics and recombination.** To characterize the dynamics of charge carrier recombination and extraction, we performed transient photovoltage/photocurrent (TPV/TPC) measurements, which offer valuable information regarding the recombination and extraction rates that follow photoexcitation by a short light pulse. A slow decay of the TPV corresponds to a large time constant for charge recombination, while a fast decay in TPC represents fast charge extraction in the device.

The TPV decay curves for CsMAFA-PbS and the control devices are shown in **Figure 5a**, showing a much slower recombination rate for the hybrid triple cation QD solar cells. The time constants $\tau_1$ and $\tau_2$ extracted from a double exponential fit (see Experimental Section) are listed as an inset table in the Figure. In particular, $\tau_2$ shows a clear enhancement: it is 50% higher than the corresponding value for $MAPbI_3$-PbS and is 200% higher than the $PbX_2$-PbS reference devices.

Transient photocurrent measurements (**Figure 5b**) show faster charge extraction within the CsMAFA-PbS films, extracted from a single exponential decay: the rate constant is approximately 50% smaller when compared to the $PbX_2$-PbS devices. The combination of these results shows that triple cation passivated PbS QD solar cells exhibit slower recombination and faster charge extraction, in agreement with the observed improvements in the photovoltaic performance of these devices.

Light intensity dependent *J-V* measurements offer further insights. **Figure 5c** shows a semilogarithmic trend in $V_{OC}$ at various light intensities *I*. The slope *S* extracted from $V_{OC} = C + S\ kT/q\ ln(I)$ (where *k* is the Boltzmann constant, *T* is the temperature and *q* is the elementary charge) allows one to extract which order of recombination is occurring. An *S* value of unity corresponds to a second order recombination (bimolecular recombination), while *S* = 2 indicates first order monomolecular or trap-assisted recombination.[37] We find the slopes to be 1.01, 1.02 and 1.04 for CsMAFA-PbS, MAPbI$_3$-PbS and PbX$_2$-PbS respectively (with an uncertainty of 0.01), suggesting that the recombination in all three types is predominantly bimolecular in nature. However, the near unity slope of the CsMAFA-PbS devices suggests that it exhibits the lowest degree of trap-mediated first order recombination,[38,39] suggesting that the CsMAFA shell is particularly effective in passivating traps in PbS QDs.

The current density at short circuit condition *versus* light intensity (**Figure 5d**) is fitted by $J_{SC}$ = A $I^\alpha$, with extracted values for $\alpha$ of 0.99, 0.97 and 0.96 for CsMAFA-PbS, MAPbI$_3$-PbS and PbX$_2$-PbS, respectively (with an uncertainty of 0.01). Deviation from $\alpha$=1 indicates the occurrence of bimolecular recombination or other intensity-dependent processes such as space-charge effects at short-circuit conditions. In all three types of devices, such effects are not significant, suggesting that all photo-generated charges are effectively extracted from the device.

These results demonstrate that triple cation coating of PbS QDs results in suppression of recombination that takes place in the device's active layer. A possible explanation for this observation can be found in the core-shell energetics of the CsMAFA-PbS QDs. For this purpose, ultraviolet photoelectron spectroscopy (UPS) spectra of CsMAFA-PbS and PbX$_2$-PbS QD solids were obtained (**Figure 5e**), from which the relative alignment of the valence bands of the core and shell, assuming a constant Fermi level, can be extracted. To estimate the energy levels of the PbS core, PbX$_2$-PbS QD films have been measured, since it is not possible to measure PbS films with no passivation at all. We note that the same estimation was recently employed by Zhang *et al.* who investigated core-shell CsPbI$_3$-PbS quantum dots.[19] We observe that the valence band of the CsMAFA shell is 0.2 eV deeper than that of the PbS core. The positions of the conduction bands were estimated using the optical gaps of the PbS core (1.3 eV) from the PL peak position measurement, and the CsMAFA perovskite shell. The latter was estimated to be 1.8 eV, corresponding to the bandgap of a 2D perovskite approximately 10-15 atoms thick, based on the HR-TEM images.[40] This estimate places the conduction band of the PbS core ~0.3 eV deeper than that of the perovskite shell. This energetic alignment picture, summarized in **Figure 5f**, suggests that the charges are effectively confined within the PbS core, suppressing recombination, yet the relatively low energy barriers cause no hinderance to charge transport between neighboring quantum dots. These observations are in agreement with the increased $J_{SC}$ and FF observed for the CsMAFA-PbS devices, as well as the larger optimal thickness for these devices.

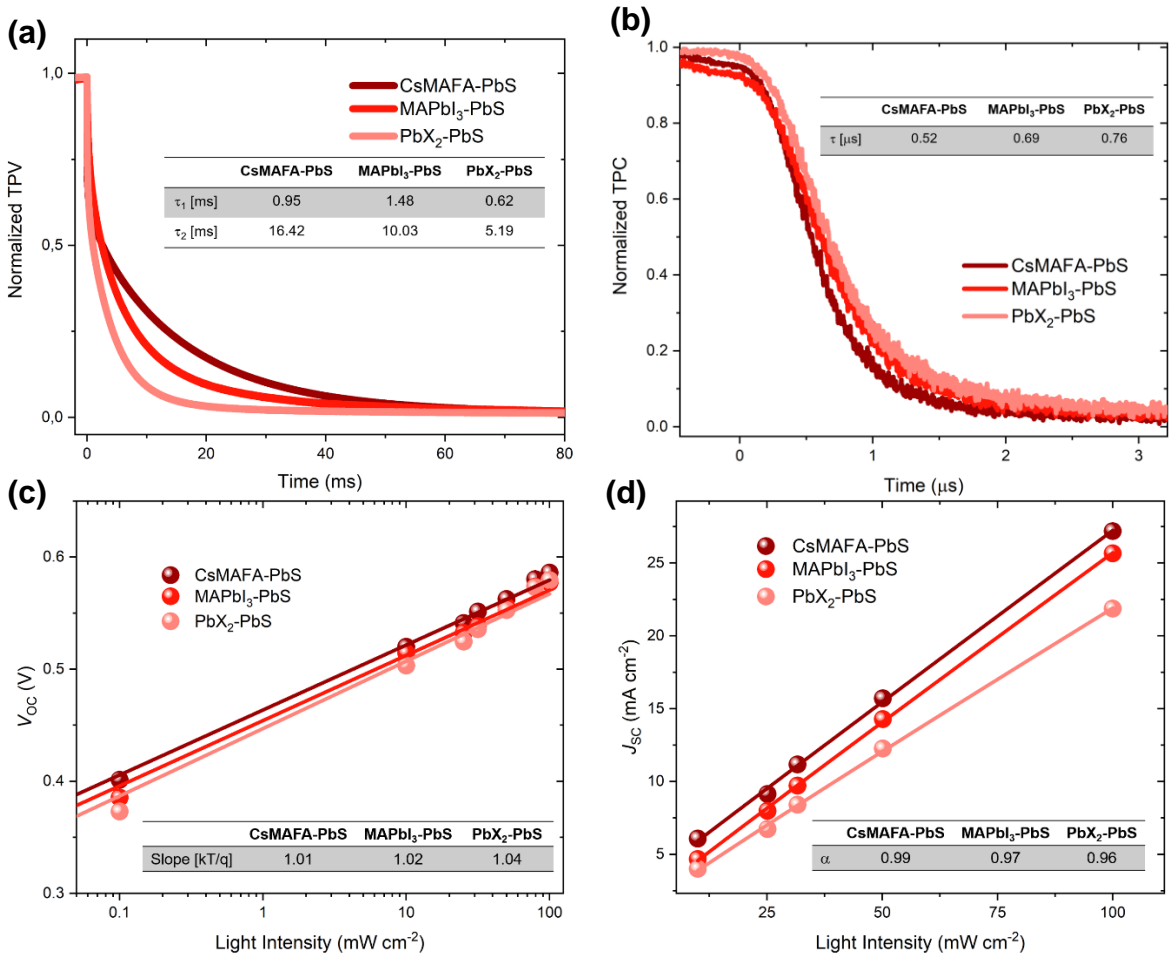

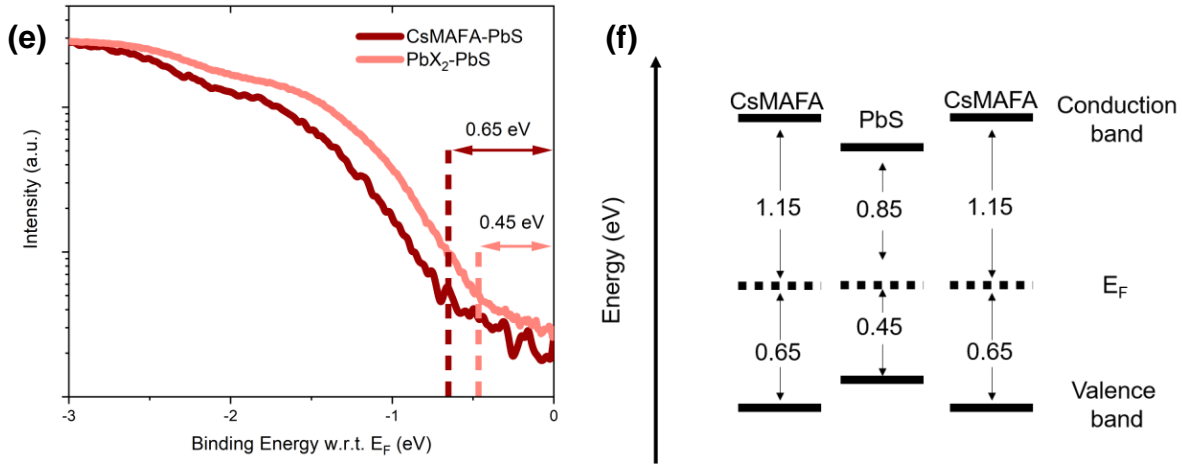

**Figure 5.** Charge carrier dynamics in the three active layer types. Transient (a) photovoltage and (b) photocurrent. Light intensity dependence of (c) $V_{OC}$ and (d) $J_{SC}$. (e) Ultraviolet photon spectroscopy (UPS) spectra, showing the position of the valence band onset with respect to the Fermi level, (f) schematic of the core and shell energy levels, extracted from UPS and optical gap measurements.

CONCLUSION

In summary, we present a solution-based process for the single step formation of triple cation perovskite coated PbS QDs. We confirm the formation of a perovskite shell by a range of spectroscopic and microscopic methods, and demonstrate that it results in higher interdot coupling and suppressed recombination. Photovoltaic devices with CsMAFA coated dots as active layers exhibit an enhanced short circuit current and fill factor, while maintaining a similar open-circuit voltage to reference devices, resulting in a maximum efficiency of 11.3%. These results surpass the performance of most previously reported hybrid perovskite-PbS solar cells, highlighting the great potential of triple cation perovskite surface passivation for efficient and stable nanocrystal-based solar-energy harvesting devices.

EXPERIMENTAL SECTION

**OA-PbS Synthesis and Triple Cation Ligand Exchange**. Oleic acid covered PbS QDs were synthesized following the hot-injection method.[26] The injection temperature was chosen to be 90°C in order to generate QDs with a bandgap around 1.3 eV. The ligand exchange starts by preparing the triple cation perovskite precursor as previously reported.[14] We create a 0.2 mol/L precursor in DMF:DMSO (4:1 volume). For the inorganic cations, the molar ratio is chosen to be 0.05:0.85:0.15 CsI:PbI$_2$:PbBr$_2$, and for the organic cations it is 5:1 FAI:MAI (stands for formamidinium iodiode and methylammonium iodide). In order to dissolve all components in the solution, we heat the inorganic precursors at 180°C for 2 min and vigorously mix afterwards. Then, 5 ml of the perovskite precursor is mixed in a centrifuge tube with 5 ml of OA-PbS (10 mg/ml in octane). After extensive vortexing, the DMF phase is collected and washed twice with octane. 5 ml of toluene is then added to the centrifuge tube, and the milky-brown suspension is centrifuged for 10 min at 4350 rpm. The supernatant is discarded, and the QDs that remain at the bottom are dried for 30 min in a vacuum chamber, and finally re-dispersed in butylamine (300 mg/ml).

**Device Fabrication**. ITO-on-glass substrates are cleaned using a sonication bath in acetone and isopropanol, followed by 10 min of oxygen plasma etching at 0.4 mbar and 100 W. Three layers of ZnO sol-gel are spin-coated on the substrate at 2500 rpm for 40 s and consecutively annealed at 100°C (first two layers, 10 min) and 300°C (last layer, 30 min). The CsMAFA-PbS active layer is spin-coated from the previously obtained dense ink at 2500 rpm for 30 s, and annealed at 100°C for 30 min. Two EDT-PbS layers are then spin coated on top to act as the hole extracting layer. The size of the EDT-PbS QDs was the same (same bandgap) as for the active layer. To form each, 20 µl of OA-PbS solution (50 mg/ml) is spin-coated at 2500 rpm for 10 s, followed by adding 80 μl EDT solution (1%vol in acetonitrile), and again spin-coated for 10 s at 2500 rpm. After 30 s, the surface is washed twice with acetonitrile. 70 nm of Au is then thermally evaporated to form the electrodes. The area of each solar cell is 4.5 mm$^2$ and there are 8 devices per substrate.

**UV-vis and FTIR Characterization**. UV-Vis spectroscopy was performed on films using glass as a substrate in a JASCO UV-Vis V670 spectrometer. FTIR spectroscopy on single side polished Si substrates was measured with a JASCO FT/IR-4600 spectrometer. All measurements are carried out in ambient air at room temperature.

**J-V Characterisation**. Current-density vs voltage sweeps were performed using a Keithley 2450 Source Measure Unit. *J-V* curves were recorded under an AM1.5G solar simulator (ABET Sun 3000 class AAA), which was corrected for spectral mismatch using a reference Si diode (NIST traceable, VLSI) and the corresponding EQE measurement. The measurements were carried out without a mask for the devices. The light intensity *JV* measurements were performed by using optical filters from Thorlabs with the corresponding optical densities.

**EQE Characterization**. The external quantum efficiency (EQE) was measured with monochromatic light from a halogen arc lamp from 350 nm to 1100 nm, using a calibration Si diode (NIST-traceable, Thorlabs).

**X-ray Photoemission Spectroscopy (XPS).** Samples were prepared on ITO-glass substrates and transferred to the ultrahigh vacuum chamber of a Thermo Scientific ESCALAB 250 Xi spectrometer. XPS experiments were performed with a XR6 monochromatic Al Kα source (hγ = 1486.6 eV) and a pass energy of 20 eV.

**X-ray Diffraction Spectroscopy (XRD).** The XRD measurements of the quantum dot films were conducted with a Rigaku SmartLab diffractometer with a 9kW rotating copper anode. 2D intensity maps were recoded using a 2D HyPix3000 detector in a coupled theta-2theta scan (beam collimator 0.5mmφ). The map was background corrected and a central profile was taken to obtain the intensity vs. 2theta diffractogram, which was normalized to account for variations in film thickness. Contributions from kβ line were stripped using the SmartLab Studio II software.

**Scanning Electron Microscopy (SEM).** SEM imaging of films and cross sections was performed using a JSM-7610F FEG-SEM (Jeol) at a chamber pressure $<10^{-6}$ mbar. Samples were mounted on standard SEM holders using conductive Ag paste to avoid sample charging. The top-view images were recorded using the secondary electron detector (LEI) at an acceleration voltage of 1.5 kV and the cross-sections using the SEI detector at an acceleration voltage of 5 kV.

**Transmission Electron Microscopy (TEM).** High-angle annular dark-field scanning TEM (HAADF-TEM) images were recorded using a FEI Tecnai 20 operated at 200 kV for high resolution imaging. The samples were prepared by drop casting 5 μl of diluted (1:10000) CsMAFA-PbS ink onto an ultrathin carbon 3 mm grid and allowed to slowly dry in air.

**Transient photovoltage/current measurement.** The TPV/TPC measurements were carried out by briefly illuminating the device with a blue LED and measuring the current/voltage traces with a Picoscope 5000 oscilloscope. The TPV/TPC curves were fitted using

$$\text{TPV} = A \exp\left(-\frac{t}{\tau_1}\right) + B \exp\left(-\frac{t}{\tau_2}\right) \quad (1)$$

$$\text{TPC} = A \exp\left(-\frac{t}{\tau}\right) \quad (2)$$

## ASSOCIATED CONTENT

**Supporting Information**.

The Supporting Information is available free of charge.

X-ray photoemission spectroscopy characterisation of CsMAFA perovskite reference sample, scanning electron microscopy image of a CsMAFA-PbS film, Photovoltaic performance evolution of CsMAFA-PbS, MAPbI$_3$-PbS and PbX$_2$-PbS photovoltaic devices.

## AUTHOR INFORMATION

**Corresponding Author**

*Prof. Dr. Yana Vaynzof, vaynzof@uni-heidelberg.de

**Author Contributions**

The manuscript was written through contributions of all authors. All authors have given approval to the final version of the manuscript. M. A. S. and D. B.-K fabricated the QDs and PV devices for this study and carried out most of the experiments. Q. S. and V. L. performed the XPS and UPS measurements. F. P. measured XRD and SEM and P. G.-O. acquired the HR-TEM images. A. D. T. and Y. V. designed and supervised the project. M. A. S, A. D. T. and Y. V. wrote the manuscript.

## ACKNOWLEDGMENT

We would like to kindly thank Prof. Uwe Bunz for providing access to the device fabrication facilities and Prof. J. Zaumseil for access to SEM. This project has received funding from the European Research Council (ERC) under the Europen Union's Horizon 2020 research and innovation programme (ERC Grant Agreement n° 714067, ENERGYMAPS).

# REFERENCES

(1) Kagan, C. R.; Lifshitz, E.; Sargent, E. H.; Talapin, D. V. Building Devices from Colloidal Quantum Dots. *Science* **2016**, *353,* 1-10.

(2) Azmi, R.; Seo, G.; Ahn, T. K.; Jang, S. Y. High-Efficiency Air-Stable Colloidal Quantum Dot Solar Cells Based on a Potassium-Doped ZnO Electron-Accepting Layer. *ACS Appl. Mater. Interfaces* **2018**, *10*, 35244-35249.

(3) Gao, J.; Luther, J. M.; Semonin, O. E.; Ellingson, R. J.; Nozik, A. J.; Beard, M. C. Quantum Dot Size Dependent J - V Characteristics in Heterojunction ZnO/PbS Quantum Dot Solar Cells. *Nano Lett.* **2011**, *11*, 1002–1008.

(4) Borchert, H. *Solar Cells Based on Colloidal Nanocrystals*; Springer International Publishing Switzerland, Basel, 2014.

(5) Lan, X.; Voznyy, O.; Kiani, A.; García De Arquer, F. P.; Abbas, A. S.; Kim, G. H.; Liu, M.; Yang, Z.; Walters, G.; Xu, J.; Yuan, M., Ning, Z., Fan, F., Kanjanaboos, P., Kramer, I., Zhitomirsky, D., Lee, P., Perelgut, A., Hoogland, S., Sargent, E. H. Passivation Using Molecular Halides Increases Quantum Dot Solar Cell Performance. *Adv. Mater.* **2016**, *28*, 299–304.

(6) Kilina, S. V.; Tamukong, P. K.; Kilin, D. S. Surface Chemistry of Semiconducting Quantum Dots: Theoretical Perspectives. *Acc. Chem. Res.* **2016**, *10*, 2127-2135.

(7) Chuang, C.-H. M.; Brown, P. R.; Bulović, V.; Bawendi, M. G. Improved Performance and Stability in Quantum Dot Solar Cells through Band Alignment Engineering. *Nat. Mater.* **2014**, *13*, 796–801.

(8) Aqoma, H.; Mubarok, M. Al; Lee, W.; Hadmojo, W. T.; Park, C.; Ahn, T. K.; Ryu, D. Y.; Jang, S. Y. Improved Processability and Efficiency of Colloidal Quantum Dot Solar Cells

TOC Figure

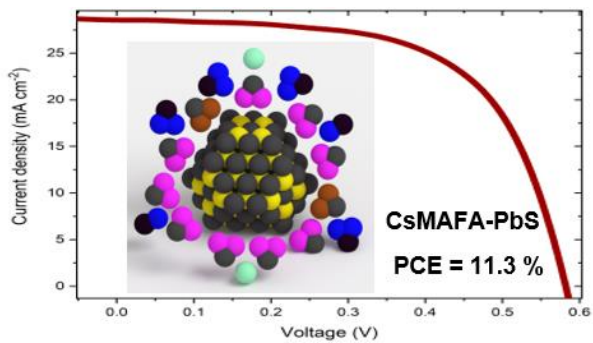